  \documentstyle[11pt,aaspp4,psfig]{article}

\def\hi{\noindent \hangindent=2.5em}
 
\begin{document}

\title{Generation of Cosmic Magnetic Fields at Recombination}

\author{Craig J.\ Hogan\altaffilmark{1}}
\affil{Departments of Astronomy and Physics, University of Washington,  Seattle,
WA  98195-1580}

\altaffiltext{1}{e-mail address: hogan@astro.washington.edu}
  
%----------------------------------------------

\begin{abstract}
It is shown that the standard cosmological model
predicts {\it ab initio} generation of   large-scale but very  
small-amplitude cosmic magnetic fields at the epoch of recombination 
of the primeval plasma. Matter velocities  dominated by coherent 
flows on a   scale $L\approx 50h^{-1}(1+z)^{-1}$ Mpc lead to a  
 dipole of radiation flux in the frame of the moving matter.  
Thomson scattering of the radiation  differentially accelerates
the electrons and ions, creating large-scale coherent   electric 
currents and magnetic fields. This process is analyzed using  
magnetohydrodynamic  equations which  include a modification of 
Ohm's law describing the effect of Thomson drag on the electrons.   
The field strength  is estimated to be  $B\simeq  10^{-20}$G. 
\end{abstract}

%----------------------------------------------

\section{Introduction}                  \label{sec:intro}
The behavior of cosmic matter and radiation at $z\approx 1100- 1300$,
the recombination era, is now broadly  understood from the 
concordance of 
direct observations (e.g. de Bernardis et al 2000, Lange et al. 2000,
Hanany et al. 2000) and
detailed theoretical models (e.g. Bond et al. 1996, 1999, Hu et al. 1996,
Lawrence et al. 1999)
of anisotropy in the cosmic background radiation. 
The amplitude and shape of the anisotropy spectrum at Legendre multipoles $l\le 400$
(and especially the first acoustic peak at $l\approx 200$)  confirm  the main
physical ingredients of the model: baryonic 
matter, thermal blackbody  radiation,  collisionless dark
matter, and large-scale primordial adiabatic perturbations. This note
analyzes the
generation of   large-scale coherent electric and  magnetic fields  in this system.
Although the fields are critical for dynamically coupling the ions and electrons
and apply forces comparable to the Thomson drag of radiation on the matter,
the stresses of the residual magnetic fields are dynamically negligible.

The process described here 
generates the fields from scratch, without the need for   an
exotic early-universe seed field: it acts as a ``battery'' rather
than a ``dynamo''. 
 The fields are generated by currents created by differential
radiation pressure on the electrons and ions as  plasma
moves under the influence of gravity; the   electrons experience
a much stronger force from 
the radiation  than the ions do, tending to create an electron-ion
drift and hence an electric  current. However, a  large-scale  electric field
arises to cancel the Thomson current, induced by the formation of a 
 large-scale magnetic field of comparable magnitude.   This effect 
makes little practical difference in 
 models of the recombination era, which assume
perfect coupling between the electrons and rest of the plasma; it is 
interesting however that the coupling actually depends on large-scale
electrical and magnetic fields, and therefore on the validity of Maxwell's
equations (including zero photon mass) over scales of about the horizon 
size (100 kpc) at
recombination.  The battery analyzed here
 resembles models of  batteries   in other
astrophysical environments (e.g.  Zeldovich et al. 1983).

\section{Currents from Thomson Drag}
  Before decoupling,   baryons and
photons are tightly coupled; the radiation pressure provides
a large restoring force   so   baryon perturbations
are oscillating acoustic waves with sound speed $c_S$ 
not much below $c/\sqrt{3}$.
Oscillations on scales with  favorable phases for maximizing
the  density perturbation  at decoupling lead to 
``acoustic peaks'' in the angular spectrum of background anisotropy.
After decoupling, the baryons no longer feel the pressure of the radiation
and their own  pressure is negligible, so they simply flow into
the dark matter potentials. We focus here on the transition epoch
around  last scattering,
when the photon path length is    larger than the scales under consideration
(so the photons are no longer tied to the baryons and oscillations 
have ceased), but when the radiation density and ionization
are  high enough that the photon drag on matter is still significant.

The generation of fields
 is controlled by  the amplitude and spatial coherence scale
of  radiation dipoles in the
matter frame, which (for small optical depth) depend mainly
on the structure of the velocity flow caused by linear perturbations.  
The 
typical rms dark matter
 velocities corresponding to   density perturbations  on
scale
$L$ are
$v_L\approx LH (\delta\rho/\rho)_L$, where $H$ is the Hubble rate and
$(\delta\rho/\rho)_L$ the rms density contrast of the dark matter. In terms
of the fluctuation power spectrum $P(k)$,   
the rms peculiar velocities $v_L\propto
k^{3/2}v_k\propto k^{1/2} P(k)^{1/2}$ in  standard
CDM are maximized at the familiar scale   determined by  the comoving
horizon size,
$ct_{eq}(1+z_{eq})= 50 h^{-1}{\rm Mpc}(\Omega_M h/0.2)^{-1}$,
 at matter-radiation equality
($1+z_{eq}=4780 [\Omega_M h/0.2]$),   and
fall  off
  (as $v_L\propto L^{-1}$ and $ L$) on larger and smaller scales.
(This is the same pattern on the same comoving scales
 as linear large-scale flows today,
but with velocities smaller by a factor $\approx (1+z)^{-1/2}$).
 
 The   dark matter velocity is
about $v/c\simeq\delta T/T\simeq 10^{-5}$ or $v\simeq 3 {\rm km\ s^{-1}}$
as each scale enters  the horizon, and in the matter-dominated era grows
thereafter as
$(1+z)^{-1/2}$. (The acoustic velocity of the baryons,
$v\approx c_S (\delta\rho/\rho)$, depends on 
the phase of the acoustic oscillations and so has  a more
complicated dependence on scale.) 
For the present discussion we adopt the simplified picture that once  the photon
path length exceeds $L_{eq}$,  the 
background radiation as viewed from the moving frame of the baryons  has a dipole
anisotropy coherent  over scales $L\approx ct_{eq}(1+z_{eq})/(1+z)$, and  
 we  adopt
$v=v_{10}\times 10{\rm km\ s^{-1}}$  as a typical value 
for baryon  velocities on this
scale.

The radiation
  dipole produces a  drag on the residual electrons 
by Thomson scattering. 
This is by far the most important dynamical interaction of the radiation
background with the matter, and remains so even after the fractional ionization
becomes small.
An electron moving  with velocity $\vec v$ relative to frame
in which the  dipole vanishes
experiences a drag 
force (e.g. Peebles 1993, Peacock 1999)
\begin{equation}
\vec F_{Thomson}=-{4\over3}\sigma_T aT_\gamma^4\vec v/c
\end{equation}
where $\sigma_T$ is the Thomson cross section and $T_\gamma=2728 z_{1000}$K is the
radiation  temperature. 
The acceleration of the electron,
 if there are no other
forces, is then
\begin{equation}
\vec a_{Thomson}=F_{Thomson}/m_e=
 -1.4\times 10^{-2} {\rm cm\ s^{-2}} z_{1000}^4 \vec v_{10}.
\end{equation}
We can ignore the corresponding direct acceleration
of  the ions by radiation; the scattering is suppressed by two powers of
mass, and the acceleration by one more.

If there are no macroscopically organized electromagnetic fields, the main other
force  experienced by an electron is friction on the ions, dominated by
long-range, small angle electron-proton scatterings out to the Debye length.
(Because of the long range of the Coulomb  force, this remains true even if
the ionization is low).  The momentum transfer between electrons
and ions of number density $n_e$ is characterised by a rate (Spitzer 1962, Shu 1992)
\begin{equation}
\nu_c \approx  3\times 10^{-3} {\rm s^{-1}} n_e T_{3000}^{-3/2}
\end{equation}
(This corresponds to an electrical resistivity
$\eta\approx c^2m_e\nu_c/4\pi n_ee^2\approx 0.6\times
10^{13}\ln\Lambda T^{-3/2}{\rm cm^2\ s^{-1}}$, with $T$ in K and
$T_{3000}=T/3000$, where in this case the Spitzer factor
$\ln\Lambda\simeq 20$.)  Thus the electron
gas, in the absence of a macroscopic  electromagnetic field, would
develop a velocity relative to the ions 
\begin{equation}
\vec v_{ie}\approx
\vec a/\nu_c=-4 {\rm cm\ s^{-1}} z_{1000}^4\vec v_{10}n_e^{-1}
T_{3000}^{3/2},
\end{equation}
in the process of transferring the  radiation drag momentum impulse to the rest of 
the plasma. In other words, this is the velocity an electron acquires
before its accumulated drift momentum is randomized by scattering.
 (Note that the
drift of charged particles relative to neutrals is much larger than this, but has no
effect on this argument and will be neglected.)
This relative velocity can develop without an accumulation of 
net charge, but corresponds to an electrical current coherent over the large   scale
of the  matter flow.\footnote{Note that because of the Thomson drag, the velocity
field   is not an irrotational potential flow, so no symmetry prevents 
current loops from forming. Proper treatment of return currents however
requires consideration of retarded fields and transients, which are
omitted here.} The
numerical value  seems like a small velocity, but the corresponding  current
density $\vec
J=-en_e\vec v_{ie}$ is actually very large. The   magnetic field on scale $L$
estimated from Ampere's law,
  $\vec\nabla\times \vec B= (4\pi/c)\vec J$,
has an amplitude
\begin{equation} \label{eqn:thomsonfield}
  B\approx  4\pi c^{-1} e n_e v_{ie} L,
\end{equation}
which for   typical numbers at recombination
yields a field with  $B\ge 10^4$ G!
This clearly violates the assumption we have made of zero fields.
It does however indicate  that  cosmic recombination 
with primordial perturbations inevitably
generates {\it ab initio}   large-scale coherent  
fields.   

\section{Magnetohydrodynamics with Thomson drag}

The field does not actually reach such a large strength; instead,
the electric field increases (mostly via induction) until the transfer of
(photon-drag-induced) momentum from the electrons to the ions and neutrals
then occurs via the field  rather than electron-ion friction.
This  reduces
  the electron-ion differential velocity, limiting the  current. 
The situation is best described with  equations of 
classical magnetohydrodynamics (Spitzer 1962, Jackson 1975, Shu 1992), but with the
addition of Thomson drag on the electrons.

The main new effect is    a 
modification of Ohm's law.  We adopt a ``laboratory frame'' 
in which the mean radiation dipole vanishes, and describe
a fluid moving with velocity $\vec v$
in which we neglect ion-neutral drift. 
The medium  responds with the
 same conductivity $\sigma=c^2/4\pi\eta$ whether
  electron accelerations arise from electric fields or from Thomson drag.
Therefore in  the fluid frame (denoted by primes),
Ohm's law for the current density  now includes a drag term, 
\begin{equation} \label{eqn:newohm}
\vec J'=\sigma[\vec E' + (m_e/e)\vec a'_{Thomson}],
\end{equation}
Transforming to the lab frame,
\begin{equation} \label{eqn:newohm}
\vec J/\sigma=\vec E + {1\over c}(\vec v\times \vec B)-
{\vec v\over c} {4\over 3} {\sigma_T aT^4\over e}.
\end{equation}
As we have seen, the  MHD approximation applies that
$\sigma$ is very large; the fields 
adjust themselves such that the  terms on the right side nearly cancel
(i.e., $\vec J/\sigma\rightarrow 0$).   (The new Thomson term however breaks
the usual MHD phenomenon of ``field freezing'', even though the conductivity is
very high.)
Therefore the electric field approximately balances
the Thomson drag everywhere,
\begin{equation} \label{eqn:electric}
\vec E\approx -{\vec v\over c} {4\over 3} {\sigma_T aT^4\over e},
\end{equation}
and has the same macroscopically organized structure
and dynamical importance---  indeed
over time the electric field does the same work on the electrons 
as the Thomson drag. 
(However  that work is almost cancelled  by work done by the ions on
the same field--- which  is the main dynamical 
coupling between ions and electrons.)

 The electric field grows quickly (due to the ``displacement current'') until
it is sufficient to cancel the Thomson drag. 
The electric field in turn grows a magnetic field by
Faraday  induction, 
\begin{equation} \label{eqn:faradaymaxwell}
\vec\nabla\times \vec E + {1\over c}{\partial \vec B\over \partial t} =0.
\end{equation}
However,  this corresponds to a very slow growth of the magnetic field; even by
the end of the process the magnetic field is only about as strong
as the electric field was all along, eq. (\ref{eqn:electric}),
  corresponding to a magnetic field of only about $10^{-20}v_{10}$G,
independent of scale (for a given $v$). 
The coherence scale of the fields is
comparable to the scale of the velocity flows, $L\approx 50h^{-1}(1+z)^{-1}$Mpc.

The  magnetic stress  is too small to affect the flow of scattering matter or 
the pattern of cosmic anisotropy.
The electron density and the radiation density 
both decrease rapidly after recombination, and  
the battery shuts down when the matter loses its
purchase on the radiation.
The Thomson drag time on
the  plasma as a whole exceeds the Hubble time  
when the ionization falls below $n_e/n\le 10^{-2} z_{1000}^{-5/2}$,
which happens for standard ionization history at $z\approx 900$.
However, the ionization remains high enough ($n_e/n\ge 10^{-3.5}$)
to keep the fields frozen to the plasma.
The fields passively follow the still nearly-uniform expanding
medium, preserving the coherent $\simeq 50h^{-1}$Mpc-scale
  comoving pattern of fields as the
  field strength  redshifts like $B\propto (1+z)^2$. 
  Linear perturbations in
the baryon density grow in the usual way, responding mainly to dark matter gravity
rather than magnetic stresses.

The magnetic fields  never become dynamically important,
although in principle they might form the seeds of dynamo fields.
In essence the process acts like other astrophysical batteries, albeit
on a much larger scale.
As  the field remains frozen to the matter, the magnetic  
field in a system of baryon density $n$ at some later time  has
\begin{equation}
\label{eqn:strength}
B\approx   10^{-22}{\rm G} (n/{\rm cm^{-3}})^{2/3}.
\end{equation} 
More likely, the observed  large scale galactic fields (see Kronberg 1994,
Beck et al. 1996 and Zweibel and Heiles 1997 for reviews),
 fully-developed microgauss galactic fields at high redshift
(Wolfe et al. 1992),     
  large-scale fields $\ge 0.2\mu$G between clusters    in 
the   intergalactic medium (Kim et al. 1989),
 and lower limits $\ge 0.1- 0.4\mu$G in the intracluster medium of galaxy clusters
(Kim et al. 1991, Rephaeli et al. 1994, Sreekumar et al. 1996) arise as an ejection of
strongly dynamo-amplified fields from  compact systems such as AGNs, and have nothing to
do with recombination.

\acknowledgements
An earlier version of this paper reached essentially opposite conclusions about
the field strength--- having wrongly guessed that the amplitude reached was
limited by equipartition rather than the induction-limited growth rate. 
I am grateful for a strong contingent of alert readers for finding
this classical error: A. Loeb, M. Rees,  D. Spergel, and especially J. Goodman.
I am grateful for critical comments by K. Jedamzik and D. Scott,
and for useful conversations with S. Phinney,  J.
Wadsley  and J. Dalcanton. This work
was supported at the University of Washington by the NSF.

%\break

\section{References}

\hi{Balbi, A. et al. 2000, ApJ submitted, astro-ph/0005124}

\hi{de Bernardis, P. et al. 2000, Nature 404, 955}

\hi{Beck, R., Brandenburg, A., Moss, D., Shukurov, A. and Sokoloff, D. 1996,
ARAA 34, 155}

\hi{Bond, J. R.,    Efstathiou, G.  and   Tegmark, M. 1997, Mon.
Not. R. Astron. Soc. 291, L33}  

\hi{Bond, J. R. and   Jaffe, A. H. 1999, Phil. Trans. R. Soc.
London 357, 57 }% astro-ph/9809043 

\hi{Farrar, G. and Piran, T. 2000, Phys.Rev.Lett. 84, 3527}

\hi{Hanany, S. et al. 2000, ApJ submitted, astro-ph/0005123}

\hi{Hu, W., Sugiyama, N. and Silk, J. 1997,  Nature 386, 37  }

\hi{Jackson, J. D. 1975,{\it Classical Electrodynamics}, Wiley}

\hi{Kamionkowski, M. and Kosowski, A. 1999,
Ann.Rev.Nucl.Part.Sci. 49, 77}

\hi{Kim, K-T., Kronberg, P. P., Giovanni, G., and Venturi, T. 1989, 
Nature 341, 720 }

\hi{Kim, K-T., Tribble, P. C., and Kronberg, P. P.,1991, ApJ 379,80 }

\hi{Kronberg, P. P. 1994, Rep. Prog. Phys. 57, 325}

\hi{Lange, A. E. et al. 2000, ApJ submitted, astro-ph/0005004}

\hi{Lawrence, C.R., Scott, D., and White, M.  1999, PASP   111, 525}

\hi{Peacock, J. 1999, {\it Cosmological Physics}, Cambridge}

\hi{Peebles, P. J. E. 1993, {\it Principles of Physical
Cosmology}, Princeton }

\hi{Rephaeli, Y., Ulmer, M. and Gruber, D. 1994, ApJ 429, 554}

\hi{Ryu, D., Kang, H. and Biermann, P. L. 1998, Astron. Astrophys. 335, 19}

\hi{Shu, F. H. 1992, {\it The Physics of Astrophysics II: Gas Dynamics},
 University Science Books}

\hi{Sreekumar, P. et al. 1996, ApJ 464, 628}

\hi{Spitzer, L. 1962, {\it The Physics of Fully Ionized Gases}, Interscience}

\hi{Waxman, E. and Loeb, A. 2000, Nature in press, astro-ph/0003447}

\hi{Wolfe, A. M., Lanzetta, K. M., and Oren, A. L. 1992, ApJ 388, 17}

\hi{Zeldovich, Ya. B., Ruzmaikin, A. A. and Sokoloff, D. D. 1983,
{\it Magnetic Fields in Astrophysics}, Gordon and Breach}

\hi{Zweibel, E. and Heiles, C. 1997, Nature 385, 131}

\end{document}